\documentclass[11pt]{article}
 \usepackage{amsmath}
 \usepackage{txfonts}
\newcommand{\be}{\begin{equation}}
\newcommand{\ee}{\end{equation}}
\newcommand{\bn}{\begin{eqnarray}}
\newcommand{\en}{\end{eqnarray}}

\def\be{\beta}

\def\der{\partial}

\begin{document}
\title{Relativistic quantum dynamics of a charged particle in cosmic string spacetime in the presence of magnetic field and scalar potential}
\author{E. R. Figueiredo Medeiros \thanks{E-mail: erfmedeiros@fisica.ufpb.br} \ 
and E. R. Bezerra de Mello \thanks{E-mail: emello@fisica.ufpb.br}
\\
Dept. de F\'{\i}sica-CCEN. Universidade Federal da Para\'{\i}ba\\
58.059-970, J. Pessoa, PB. C. Postal 5.008. Brazil}
\maketitle
\begin{abstract}
In this paper we analyze the relativistic quantum motion of charged spin$-0$ and spin$-\frac{1}{2}$ particles in the presence of a uniform magnetic field and scalar potentials in the cosmic string spacetime. In order to develop this analysis, we assume that the magnetic field is parallel to the string and the scalar potentials present a cylindrical symmetry with their center on the string. Two distinct configurations for the scalar potential, $S(r)$, are considered: $(i)$ the potential proportional to the inverse of the polar distance, i.e.,  $S\propto1/r$, and $(ii)$ the potential proportional to this distance, i.e., $S\propto r$. The energy spectra are explicitly computed for different physical situations and their dependences on the magnetic field strength and scalar coupling constants are presented. 
\\PACS numbers: $03.65.Pm, 03.65.Ge$ 
\end{abstract}

\renewcommand{\thesection}{\arabic{section}.}

\section{Introduction}

The analysis of a quantum mechanical system under the influence of the gravitational field has attracted attention in particle physics since many years ago. In this direction, the hydrogen atom in curved spacetime has been investigated by \cite{AS,Parker,Parker1}. It was shown in \cite{Parker} that the shifts in the energy spectrum caused by local curvature is different from the usual gravitational Doppler shift. However, this shift is appreciable only in the region of strong gravitational field. Recently, the analysis of the influence of the spacetime produced by topological defects on the energy spectrum of the hydrogen atom, has been considered under both non-relativistic \cite{GB} and relativistic \cite{GB1} points of view. In these papers, the hydrogen atom is placed in the spacetime produced by an idealized linear cosmic string and a point-like global monopole. Different from the previous analysis, due to the specific geometry associated with the spacetime produced by these idealized topological defects, in these more recent investigations the energy spectra associated with the hydrogen atom can be exactly obtained.

According to the modern concepts of theoretical physics, different types of topological defects may be produced by the vacuum phase transition in the early Universe \cite{Kibble,Vilenkin}. These include domain walls, cosmic strings and monopoles. Among them, cosmic strings and monopoles seem to be the best candidates to be observed. The former are linear defects, and the spacetime produced by an idealized cosmic string is locally flat, however globally conical, with a planar angle deficit determined by the string tension. Due to this conical structure, a charged particle placed at rest in the cosmic string spacetime becomes subjected to a repulsive electrostatic self-interaction \cite{Linet1,Smith} proportional to the inverse of the polar distance from the string. Also it was shown in \cite{Furtado} that linear electric and magnetic sources in the spacetime of a cosmic string become subject to induced self-interactions.\footnote{These induced self-interactions are consequence of a distortion on the electric and magnetic fields caused by the planar angle deficit produced by the defect.} 

The non relativistic quantum mechanical analysis of the motion of a charged particle in the cosmic string spacetime taking into account the electrostatic self-interaction and in the presence of a uniform magnetic field parallel to the string has been developed in \cite{Fur}. There it was shown that the presence of the defect reduces the degeneracy degree of the Landau levels. In addition, the inclusion of the electrostatic self-interaction in this analysis breaks the residual degeneracy. In both situations, exact wave functions and energy eigenvalues have been found for specific magnitudes of the magnetic field. Moreover, in \cite{Mello} the analysis of the non-relativistic motion of two charged particles on a cone and in the presence of a static magnetic field has been developed, too.

The well-known procedure to introduce the coupling between a charged particle and electromagnetic fields in the Dirac and Klein-Gordon equations, is through the minimal coupling. By this procedure, the four-vector differential operator $p_\mu=i\partial_\mu$ is modified in order to include the electromagnetic four-vector, $A_\mu$, as $p_\mu \longrightarrow p_\mu-eA_\mu$. Some years ago, Dosch, Jansen and M\"uller in \cite{DJM}, pointed out that the minimal coupling is not the only way to couple a potential to the Dirac equation. There, it was suggested that a non-electromagnetic potential can be taken into account by making a modification in the mass term as $M\longrightarrow M+S({\vec{r}},t)$, being $S({\vec{r}},t)$ the scalar potential. This new formalism has been used by Soff, M\"uller, Rafelski and Greiner in \cite{GBW}, to analyze the Dirac equation in the presence of a Coulomb potential and a static scalar potential proportional to the inverse of the radial distance. More recently, the electrostatic self-interaction on a charged particle in the cosmic string spacetime has been considered in \cite{Nail} and \cite{Spi}, as a scalar potential to analyze its relativistic quantum motion. So, if one wants to investigate the relativistic quantum motion of a charged particle in the presence of electromagnetic and scalar potentials, both procedures, the minimal coupling and a modification in the mass term, should be taken into account. 

In this paper we shall analyze the relativistic quantum motion of bosonic and fermionic massive charged particles in the spacetime surrounding an idealized cosmic string in the presence of a uniform magnetic field parallel to the string and also in the presence of scalar potentials. Two specific configurations to the scalar potential will be considered: $(i)$ the potential proportional to the inverse of the polar distance to the string, $S \propto 1/r$, and $(ii)$ the potential linear to the polar distance to the string, $S \propto r$. We shall see that, for particular values of the magnetic field, we are able to obtain exact solutions to the Klein-Gordon and Dirac equations, providing the energy spectra associated with bound states.

Spherically symmetric linear scalar potential has been introduced into the Dirac equation in \cite{Chritc} as a simple model to confine relativistic quark states. In this paper some analytical results could be compared with those provide by the {\it bag} model. In addition to the linear scalar potential a scalar Coulomb-like potential has also been included in \cite{Chritc1}. In the latter, separete analysis of the energy spectrum are examined for the case of absence of the confining potential, followed by the complete analysis, i.e., taking into account the linear potential. In these two papers, due to the presence of linear potential, only approximated solutions for the energy spectra have been obtained. The spectra found are similar to the harmonic oscilator. Moreover, exact solutions to the Dirac equation in presence of an uniform electric field and a linear scalar potential have been found in \cite{Su}. In the latter the zero-component of the vector potential and the linear scalar potential are both proportional to the $z-$coordinate. 

This paper is organized as follows: In section \ref{sec2} we introduce the physical scenario, namely the explicit expressions for the metric tensor, for the four-vector potential and for the scalar potentials. We consider these background fields in the analyses of relativistic quantum motion of charged bosonic and fermionic particles. The first part of this paper deals with the bosonic case. In section \ref{bosonfields}, the modified Klein-Gordon equation is presented. Considering stationary states, $\Psi({\vec{r}},t)=e^{-iEt}\Psi_E({\vec{r}})$, and admitting for $\Psi_E({\vec{r}})$ an \textit{ansatz} that obeys a cylindrical symmetry, we provide, in subsection \ref{sec3}, the most general differential equation obeyed by the radial function. The obtainment of solutions for this differential equation is considered in section \ref{sec4}, specifying separately  in each subsection, different values attributed to the parameters associated with the magnetic field and scalar coupling constants. In the second part, we repeat this analysis for the fermionic case. In section \ref{fermionfields}, we present the modified Dirac equation, and our choice for the curved Dirac matrices. The following section deals with the radial differential equations resulting after the consideration of an \textit{ansatz} suggested by the symmetry of the system. Solutions to these equations and physical informations are obtained in section \ref{soldirac}. In section \ref{concl} we summarize all results. Finally in the appendix \ref{app} we present some properties associated with the most relevant differential equation presented in this paper, that is the biconfluent Heun (BCH) differential equation. Here we use natural units, $\hbar=c=G=1$.

\section{The System}
\label{sec2}

In this section we present the physical environment under which we shall consider relativistic charged particles. We first introduce the spacetime background of an idealized cosmic string where the particle will propagate, followed by vector potential associated with a uniform magnetic field parallel to the string and the most general scalar potential which allow us to obtain analytical solutions of the corresponding Klein-Gordon and Dirac equations. 

By using cylindrical coordinates with the cosmic string along the $z-$axis, the corresponding metric tensor is defined by the line element below:
\bn
	ds^2=dt^2-dr^2-\alpha^2r^2d\phi^2-dz^2 \ , \label{cs1}
\en
where the polar coordinate, $r\geq 0$, the coordinates $(t, \ z) \in(-\infty,  \infty)$ and the angular variable $\phi\in[0, \ 2\pi]$. The parameter $\alpha$, smaller than unity, is given in terms of the linear mass density of the string, $\mu$, by $\alpha=1-4\mu$.

In the cosmic string spacetime, the vector potential associated with an uniform magnetic field parallel to the string, $\vec{B}=\vec{\nabla}\times\vec{A}=B_0 \ \hat{k}$, in Coulomb gauge can be expressed by,
\bn
\label{Mag}
\vec{A}=(0, \ A_\phi, \ 0), \quad {\rm with} \quad A_\phi=-\frac12\alpha B_0r^2 \ .
\en

Also we shall consider in this analysis the presence of cylindrically symmetric scalar potentials which allows us to provide analytical solutions for the field equations. This scalar potential reads:\footnote{The relativistic quantum analysis of the motion of a charged spin$-0$ particle in the spacetime produced by an idealized cosmic string and in the presence of magnetic field produced by a magnetic monopole and scalar potentials were developed in \cite{ALO}.}
\bn
\label{potential}
S(r)=\frac{\eta_C}r+\eta_L r \ .
\en

In the next sections we shall develop the relativistic quantum mechanical analysis associated with spin$-0$ and spin$-1/2$ charged particles under the influence of the above field configurations. We shall start with the bosonic case, followed by the fermionic one.

\section{Bosonic Fields: Klein-Gordon Equation}
\label{bosonfields}

The relativistic quantum motion of a charged bosonic particle in curved space, and in the presence of electromagnetic and scalar potentials, is described by the modified Klein-Gordon equation below,
\bn
\label{KG}
	\left[{\cal D}^2-\xi R+\left(M+S(r)\right)^2\right]\Phi(x)=0 \ ,
\en
where the differential operator above is given by
\bn
\label{D}
	{\cal D}^2=\frac{1}{\sqrt{-g}}D_\mu\left(\sqrt{-g}\,g^{\mu\nu}D_\nu\right) \ ,
\en
with $D_\mu=\der_\mu+ieA_\mu$, $g={\rm det}(g_{\mu\nu})$ and $S(r)$ is the scalar potential. Since we are considering a minimal coupling, the previously introduced non-minimal coupling $\xi$ needs to be discarded. 

\subsection{Bosonic Radial Equation}\label{bosonradeq}
\label{sec3}

Here in this subsection we shall construct the radial differential equation associated with the Klein-Gordon equation (\ref{KG}), considering the most general physical situation involving the presence of the magnetic field and the scalar potentials. 

Due to the cylindrical symmetry of such system, let us adopt for the wave function the {\it ansatz} below:
\bn
\label{Phi}
\Phi(x)=e^{i(kz+m\phi-Et)}R(r) \ .
\en
Substituting (\ref{Phi}) into (\ref{KG}), we find that the radial function, $R(r)$, must obey the differential equation
\bn
\label{radeq}
\left[\frac{d^2}{dr^2}+\frac1r\frac{d}{dr}\right]R(r)+\left[E^2-k^2+2M\omega \frac{m}\alpha-M^2\omega^2r^2 -\frac{m^2}{\alpha^2r^2} -\left(M+S(r)\right)^2 \right]R(r)=0 \ ,
\en
where 
\bn
\omega = \frac{eB_0}{2M} \ ,
\en
is the so called \textit{cyclotron frequency} of the particle. Considering the scalar potential (\ref{potential}), the equation (\ref{radeq}) is expressed as a biconfluent Heun differential equation \cite{Heun}. 

In order to solve the above differential equation, it is convenient to define a new radial function, $u(r)$, as shown below:
\bn
\label{newradial}
u(r) = \sqrt{r}R(r) \ .
\en
Taking the scalar potential (\ref{potential}) and defining a new dimensionless variable
\bn
x = \sqrt{\Delta}r  \quad {\rm with} \quad \Delta=\sqrt{\eta_L^2 + M^2\omega^2} \ ,
\en
the radial biconfluent Heun equation \eqref{radeq} takes on the following Schr\"odinger form
\bn
\label{radialgen}
\frac{d^2u(x)}{dx^2} + \left[\frac{\epsilon_{k,m}}{\Delta} + \frac{\gamma_m}{x^2} + \frac{\delta_C}{x} + \frac{\delta_L x}{\Delta} - x^2\right]u(x) = 0\ .
\en
where
\bn
\epsilon_{k,m} &=& E^2 - k^2 - M^2 + 2\left(M\omega \frac{m}{\alpha} - \eta_C\eta_L \right) \ ,\\
\gamma_m &=& \frac{1}{4} - \eta_C^2 - \frac{m^2}{\alpha^2} \ , \\
\delta_L &=& -\frac{2M}{\sqrt{\Delta}}\eta_L \quad  {\rm and} \quad \delta_C = -\frac{2M}{\sqrt{\Delta}}\eta_C \ .
\en

After we have constructed the complete radial differential equation, the next steps are to obtain the complete wave functions set and the corresponding energy spectra.

\section{Analytical Solutions for the Klein-Gordon Equation}
\label{sec4}

In this section, we shall develop the analysis of the radial differential equation obtained in the previous section considering different physical situations. For completeness only, we start with the relativistic version of the Landau levels for bosonic charged particles. For this case the solutions for the radial differential equation are presented in terms of the confluent hypergeometric function. The newest and most relevant analysis about the relativistic quantum motion contained in this paper are exhibited in the following subsections. There we shall see that the solutions for the radial equation are given in terms of biconfluent Heun functions.

\subsection{Landau Levels}
\label{sub1}

The simplest particular solution to (\ref{radialgen}) can be found by admitting $\eta_C = \eta_L = 0$ but keeping $\omega \neq 0$. As we know, only bound states exist in such condition. The differential equation \eqref{radialgen} can be rewritten as follows:
\bn
\frac{d^2u}{dx^2}+\left[\frac{E^2-k^2-M^2}{M\omega} + \frac{2m}{\alpha} + \frac{\frac{1}{4} - \frac{m^2}{\alpha^2}}{x^2} - x^2\right]u=0.
\en
It is helpful to analyze the asymptotic behavior of solutions to the equation above, for $x\longrightarrow 0$ and $x\longrightarrow \infty$. This analysis indicates that it is useful to write $u(x)$ in the form
\bn\label{ans}
u(x) = x^{\tilde{\beta}}e^{-\frac{x^2}{2}}F(x),
\en
where $\tilde{\beta} = \frac{1}{2}+ \frac{|m|}{\alpha}$, and $F(x)$ is an auxiliary function. In fact, substituting \eqref{ans} into the differential equation we get,
\bn
\label{eq17}
x\frac{d^2F}{dx^2} + \left(2\tilde{\beta}-2x^2\right)\frac{dF}{dx}+ \left(\frac{E^2 - k^2 - M^2}{M\omega} + \frac{2m}{\alpha} - 2\tilde{\beta}-1\right)xF=0 \ .
\en
Defining a dimensionless variable $y=x^2 = M\omega r^2$, the above equation yields
\bn\label{difeq1}
y\frac{d^2F(y)}{dy^2} + (c-y)\frac{dF(y)}{dy} - aF(y) = 0 \ ,
\en
with the constants being given by
\bn
c = \frac{|m|}{\alpha} + 1 \ , \ a = \frac{1}{2}\left(\frac{|m| - m}{\alpha} + 1\right) - \frac{E^2-k^2-M^2}{4M\omega}  \ .
\en
The solution to the differential equation \eqref{difeq1} is the well-known confluent hypergeometric function:
\bn
F(y) = {}_1F_1\left(\frac{1}{2}\left(\frac{|m| - m}{\alpha} + 1\right) - \frac{E^2-k^2-M^2}{4M\omega} ,\quad \frac{|m|}{\alpha} + 1;\quad y\right).
\en
The asymptotic behavior of a confluent hypergeometric function for large values of its argument is \cite{Abra}:
\bn
{}_1F_1(a, \ b, \ z)\approx\frac{\Gamma(b)}{\Gamma(a)}\ e^z \ z^{a-b}\left[1+O(|z|^{-1})\right] \ .
\en

So, due to the divergent behavior of the function $F(y)$ for large values of its argument, bound states solutions can only be obtained by imposing that this function becomes a polynomial of degree $n$. In this case, the radial solution presents an acceptable behavior at infinity. This condition is obtained by
\bn
\frac{1}{2}\left(\frac{|m| - m}{\alpha}+1\right) - \frac{E^2 - k^2 - M^2}{4M\omega} = -n, \quad n = 0, 1, 2, 3...
\en
This equation provides the quantization condition on the energy spectrum of the particle:
\bn
\label{L-L}
E_{k, m, n} = \pm \left\{k^2 + M^2 + 2M\omega\left[2n+1+\frac{|m| - m}{\alpha}\right]\right\}^{\frac{1}{2}}.
\en
The presence of the factor $\alpha$ visibly modifies the degenerate spectrum of the particle. Each set of wave functions with the same value of $n$ is called a \textit{Landau level}. In the case of $\alpha=1$, the non-relativistic analyses of Landau levels is most developed by using Cartesian coordinates, and the spectrum can be find in many textbooks. However, by using cylindrical coordinates system, the non-relativistic analysis has been developed in \cite{Landau} as a separate problem and also in \cite{Cooper}; as to the relativistic approach this problem has been considered in \cite{Sokolov}. In \cite{Landau} the approach adopted to find the energy spectrum is similar to ours; there it is shown that the energy depends on the angular quantum number in the same combination as presented in the equation above.\footnote{Also in \cite{Fur} the energy levels depend on the angular quantum number in the same manner.}

\subsection{Vanishing magnetic field}
\label{sub2}

Let us now admit that $\omega=0$, but take into account the effects of the scalar potential, i.e., keeping non-vanishing $\eta_C$ and $\eta_L$. In this case we can simply rewrite equation \eqref{radialgen} as follows:\footnote{The case with $\omega=\eta_L=0$ has been analyzed in \cite{Nail}.}
\begin{align}
\label{radialgen2}
&\frac{d^2u(x)}{dx^2} + \left[\frac{\bar{\epsilon}_{k}}{|\eta_L|} + \frac{\gamma_m}{x^2} + \frac{\bar{\delta_C}}{x} + \frac{\bar{\delta_L} x}{|\eta_L|} - x^2\right]u(x) = 0,
\end{align}
with
\bn
\bar{\epsilon}_{k}&=& E^2 - k^2 - M^2 - 2\eta_C\eta_L \ , \nonumber \\
\bar{\delta}_L &=& -\frac{2M}{\sqrt{|\eta_L|}}\eta_L \quad  {\rm and} \quad \bar{\delta}_C = -\frac{2M}{\sqrt{|\eta_L|}}\eta_C \ .
\en
In order to obtain the energy spectrum to equation \eqref{radialgen2}, it is convenient to analyze its asymptotic behavior for $x\longrightarrow 0$ and $x\longrightarrow \infty$. Through this analysis it is possible to express the function $u(x)$ in terms of the unknown function $F(x)$ as shown below:
\bn
\label{ansatz2}
u(x) = x^\beta e^{-\frac{1}{2}x\left(x - \frac{\bar{\delta}_L}{|\eta_L|}\right)}F(x) \ ,
\en
with
\bn
\beta= \frac{1}{2}+\sqrt{\eta_C^2 + \frac{m^2}{\alpha^2}} \ .
\en
Now, substituting the ansatz \eqref{ansatz2} into \eqref{radialgen2}, the resulting equation reads:
\bn
\label{Heun2}
xF''(x) + \left(2\beta + \frac{\bar{\delta}_L}{|\eta_L|}x - 2x^2\right)F'(x) + \left[\beta\frac{\bar{\delta}_L}{|\eta_L|} + \bar{\delta}_C + \left(\frac{\bar{\epsilon}_k}{|\eta_L|} + \frac{\bar{\delta}_L^2}{4\eta_L^2}-2\beta - 1\right)x\right]F(x) = 0 \ .
\en

This equation is the biconfluent Heun's differential equation \cite{Heun}, whose solution is the so called biconfluent Heun (BCH) function, $H_B$:
\bn
F(x) = H_B\left(2\beta -1,\quad\frac{\bar{\delta}_L}{|\eta_L|}, \quad\frac{\bar{\epsilon}_k}{|\eta_L|} + \frac{\bar{\delta}_L^2}{4\eta_L^2}, \quad 2\bar{\delta}_C, \quad -x \right) \ .
\en
We shall use the Frobenius method to obtain solutions for \eqref{Heun2}:
\bn
\label{Frob}
F(x)=\sum_{n=0}^\infty \ c_n x^n \ .
\en
Substituting the series above into \eqref{Heun2}, the coefficients are defined by a three-term recurrence relation,
\bn
\label{rec}
c_{n+2}= \frac{1}{(n+2)(2\beta + n + 1)}\left[\left(-\frac{\bar{\delta}_L}{|\eta_L|}(n+1) - \bar{B}\right)c_{n+1} + (2n-\bar{C})c_n\right]
\en
and by the relation below,
\bn
c_1 = \frac{M}{\sqrt{|\eta_L|}}\left(\frac{\eta_L}{|\eta_L|}+\frac{\eta_C}{\beta}\right)c_0 \ ,
\en
where $\bar{B} = \beta\frac{\bar{\delta}_L}{|\eta_L|} + \bar{\delta}_C$ and $\bar{C} = \frac{\bar{\epsilon}_{k,m}}{|\eta_L|} - 2\beta - 1 + \frac{M^2}{|\eta_L|}$. 

A special kind of exact solutions representing bound states, can be obtained by searching for polynomial expressions to $F(x)$. Being $n$ the order of such polynomial, two different conditions must be simultaneously satisfied: 
\bn
\label{cond1}
\bar{C} = 2n
\en
being $n$ a positive integer number, and
\bn
\label{cond2}
c_{n+1}= 0 \ .
\en
Admitting that $c_0=1$ and using the recurrence relation (\ref{rec}) we present below the first coefficients for the Frobenius series:
\begin{eqnarray}
c_2&=&\frac{1}{2(2\beta + 1)}\left[\frac{2M^2}{|\eta_L|}\left(\frac{\eta_L}{|\eta_L|}+\frac{\eta_C}{\beta}\right)\left(\eta_C + \frac{\eta_L}{|\eta_L|}(\beta + 1)\right) - \bar{C}\right] \ , \nonumber\\
c_3&=&\frac{M}{6(\beta + 1)\sqrt{|\eta_L|}}\left\{\frac1{2\beta+1}\left[\frac{\eta_L}{|\eta_L|}(\beta + 2) + \eta_C\right]\left[\frac{2M^2}{|\eta_L|}\left(\frac{\eta_L}{|\eta_L|}+\frac{\eta_C}{\beta}\right)\left(\eta_C+\frac{\eta_L}{|\eta_L|}(\beta + 1)\right) - \bar{C}\right]\right.\nonumber\\
& +&\left. \left(2 - \bar{C}\right)\left(\frac{\eta_L}{|\eta_L|}+\frac{\eta_C}{\beta}\right)\right\} \ .
\end{eqnarray}

The condition $\bar{C} = 2n$ provides the following energy spectrum for the particle:
\bn
E_{k, m, n} = \pm\left[k^2 + 2\eta_C\eta_L + 2|\eta_L|\left(n+\beta+\frac{1}{2}\right)\right]^{\frac{1}{2}}.
\en
This expression holds for both signs, and the quantity inside the brackets is clearly positive if $\eta_L$ and $\eta_C$ have the same sign, so the system allows positive and negative solutions for the energy spectrum. At first glance, this equation indicates that the energy spectrum does not depend on the mass of the particle. However, we would like to mention that that the condition $c_{n+1}=0$, provides an algebraic expression involving specific values of the scalar potential coupling constant, the mass of the particle and angular momentum quantum number. Let us assume that the parameter $\eta_L$ can be adjusted to fit this equation. In this case, this parameter depends on the quantum numbers $n$ and $m$, which we define now by $\eta_{L_{n,m}}$.

In what follows, we shall consider for the function $F(x)$ the simplest case only, a polynomial of first order. So, after a few steps we can write:
\bn
F_{1,m}(r) = 1+ M \left(1+ \frac{\eta_C}{\beta}\right)r,
\en
Also for this specific case the energy reads,
\begin{align}
&E_{k,m,1}=\pm\left[k^2 + 2\eta_C\eta_{L_{1,m}} + 2|\eta_{L_{1,m}}|\left(\beta+\frac{3}{2}\right)\right]^{\frac{1}{2}}.
\end{align}
Moreover, the condition \eqref{cond2} allows us to express the parameter $\eta_{L_{1,m}}$ in terms of the mass of the particle as shown below:
\bn
\eta_{L_{1,m}} = \frac{M^2}{\beta}\left[\eta_C^2 + \left(2\beta+1\right)\eta_C + \beta(\beta + 1)\right] \ .
\en
So the corresponding energy is completely defined, and we can say that the energy spectrum of the particle, in fact, \textit{does} depend on its mass, as should be. Moreover, the energy spectrum also depends on the parameter $\alpha$ through the definition of $\beta$.

\subsection{Linear Confinement}
\label{sub3}

In this subsection we shall analyze the situation where the the Coulomb-type scalar interaction is absent, but keeping the linear confining potential and a magnetic field. For this system, the function $u(x)$ satisfies the differential equation below, obtained directly from \eqref{radialgen}:
\bn
\label{radialgen3}
\frac{d^2u(x)}{dx^2} + \left[\frac{\tilde{\epsilon}_{k,m}}{\Delta} + \frac{\tilde{\gamma}_m}{x^2} + \frac{\delta_L x}{\Delta} - x^2\right]u(x) = 0 \ ,
\en
where
\bn
\tilde{\epsilon}_{k,m} = E^2 - k^2 - M^2 + 2M\omega \frac{m}{\alpha}  \quad {\rm and} \quad \tilde{\gamma}_m = \frac{1}{4}- \frac{m^2}{\alpha^2} \ .
\en
Again, analyzing the asymptotic behavior of the differential equation for large and small values of the variable we verify that $u(x)$ can be expressed by means of an unknown function $G(x)$ as follows:
\bn
\label{uG}
u(x) = x^{\tilde{\beta}} e^{-\frac{1}{2}x\left(x - \frac{\delta_L}{\Delta}\right)}G(x) \ ,
\en
where
\bn
\tilde{\beta}= \frac{1}{2}+ \frac{|m|}{\alpha} \ .
\en
Substituting \eqref{uG} into \eqref{radialgen3}, we can see that $G(x)$ satisfies the BCH differential equation with one of its parameters being zero:
\bn\label{BCH2}
x\ G''(x) + \left(2\tilde{\beta} + \frac{\delta_L}{\Delta}x - 2x^2\right)G'(x) + \left[\tilde{\beta}\frac{\delta_L}{\Delta} + \left(\frac{\tilde{\epsilon}}{\Delta} + \frac{\delta_L^2}{4\Delta^2} - 2\tilde{\beta} - 1\right)x\right]G(x)=0 \ .
\en
Consequently, we can write
\bn
G(x) = H_B\left[2\frac{|m|}{\alpha},\quad \frac{\delta_L}{\Delta},\quad \frac{\tilde{\epsilon}}{\Delta} + \frac{\delta_L^2}{4\Delta^2},\quad 0,\quad -x\right] \ .
\en
Again, using the Frobenius expansion,
\bn
G(x)=\sum_{n=0}^\infty \ d_n\ x^n \ ,
\en
into \eqref{BCH2}, we find that the coefficients of this series obey the following recurrence relations:
\bn
\label{recur2}
d_{n+2} = \frac{1}{(n+2)(2\tilde{\beta} + n + 1)}\left[-\frac{\delta_L}{\Delta}\left(\tilde{\beta} +n+1\right)d_{n+1} + (2n-\tilde{C})d_n\right] \quad {\rm and}\quad d_1 = -\frac{\delta_L}{2\Delta}d_0 \ ,
\en
where ${\tilde{C}}= \frac{\tilde{\epsilon}}{\Delta} - 2\tilde{\beta} - 1 + \frac{\delta_L^2}{4\Delta^2}$, and we have set $d_0$ equal to unity. Below we provide two additional coefficients:
\begin{align}
d_2&=\frac{1}{2\tilde{\beta} + 1}\left[\frac{\eta_L^2}{M\bar{\omega}^3}\left(\tilde{\beta} + 1\right) - \frac{\tilde{C}}{2}\right] \ , \nonumber\\
d_3&=  \frac{1}{6(\tilde{\beta} + 1)}\frac{\eta_L}{\bar{\omega}\sqrt{M\bar{\omega}}}\left\{\frac{2(\tilde{\beta} + 2)}{2\tilde{\beta} + 1}\left[\frac{\eta_L^2}{M\bar{\omega}^3}(\tilde{\beta} + 1)-\frac{\tilde{C}}{2}\right]+ 2 - \tilde{C}\right\} \ .
\end{align}

Once again, here we are looking for polynomial solutions for the function $G(x)$. In order to do that we must impose conditions equivalent to \eqref{cond1} and \eqref{cond2} on the coefficients.

From the condition $\tilde{C}=2n$, we obtain the energy spectrum:
\bn
E_{k,m,n} = \pm\left[k^2 +\frac{ M^4\omega^2}{\eta^2_L+M^2\omega^2} - \frac{2m}{\alpha}M\omega + 2M\bar{\omega}\left(n+\frac{|m|}{\alpha} + 1\right)\right]^{\frac{1}{2}}.
\en
with $\bar{\omega} = \frac{1}{M}\sqrt{\eta_L^2 + M^2\omega^2}$. Like in the previous analysis, we can see that the energy spectrum of the particle depends on a given set of parameters. The condition $d_{n+1}=0$ provides an algebraic equation involving this set of parameters and the mass of the particle. Here, in this subsection, we shall adopt a different procedure as we did in the last subsection. We shall admit that the strength of the magnetic field is adjustable according this equation which in its turn will depend on the quantum numbers $n$ and $m$. This will provide discrete values for $B_0$, and consequently for $\omega$ which we define now by $\omega_{n,m}$.

To illustrate, let us apply this formalism for the case where the function $G(x)$ is a polynomial of first order. In this case the magnitude of the magnetic field will be given by $\bar{\omega}_{1,m}$ as exhibited in \eqref{param2}. The function $G_1(r)$ reads,
\bn
G_{1,m}(r) = 1 + \frac{\eta_L}{\bar{\omega}_{1,m}}r \ .
\en
As to the self-energy we have:
\bn
E_{k,m,1} = \pm\left[k^2 +\frac{ M^4\omega_{1,m}^2}{\eta^2_L+M^2\omega_{1,m}^2}- \frac{2m}{\alpha}M\omega_{1,m} + 2M\bar{\omega}_{1,m}\left(\frac{|m|}{\alpha} + 2\right) \right]^{\frac{1}{2}}.
\en
with the parameter $\bar{\omega}_{1,m}$ assuming the form
\bn\label{param2}
\bar{\omega}_{1,m} = \left[\frac{\eta_L^2}{M}\left(\frac{|m|}{\alpha} + \frac{3}{2}\right)\right]^{\frac{1}{3}} \ .
\en
For any values of the physical quantities, equation \eqref{param2} presents at least one real physical solution for $B_0$.

\subsection{General Solution}
\label{sub4}

In this subsection we shall investigate the solution of the problem under consideration in its more general form, which is the analysis of the quantum motion of a bosonic charged particle in the presence of a uniform magnetic field, given in terms of a vector potential by \eqref{Mag} and a scalar potential given by \eqref{potential} in the cosmic string spacetime. The corresponding radial differential equation associated with this problem is \eqref{radialgen}.

As we did previously, in order to obtain the energy spectrum, it is convenient to analyze the asymptotic behavior of \eqref{radialgen} for $x\longrightarrow 0$ and $x\longrightarrow \infty$. After we have made this analysis we can express the function $u(x)$ as follows:
\bn
u(x) = x^\beta e^{-\frac{1}{2}x\left(x - \frac{\delta_L}{\Delta}\right)}H(x) \ ,
\en
with
\bn
\beta= \frac{1}{2}+\sqrt{\eta_C^2 + \frac{m^2}{\alpha^2}} \ .
\en
\label{eq55}
Substituting the above function into \eqref{radialgen}, we get,
\bn
xH'' + \left[2\beta + \frac{\delta_L}{\Delta}x - 2x^2\right]H' + \left[\beta\frac{\delta_L}{\Delta}+\delta_C + \left(\frac{\epsilon}{\Delta} + \frac{\delta_L^2}{4\Delta^2}-2\beta-1\right)x\right]H=0 \ .
\en
Once more, by using the Frobenius method,
\bn
\label{H-func}
H(x)=\sum_{n=0}^\infty c_n \ x^n \ ,
\en
we can observe that the recurrence relations obeyed by the coefficients of this expansion is:
\bn
\label{coeff1}
c_{n+2}& =& \frac{1}{(n+2)(2\beta + n + 1)}\left[\left(-\frac{\delta_L}{\Delta}(n+1) - B\right)c_{n+1} + (2n-C)c_n\right] \ , \nonumber\\
c_1& =&-\frac{1}{2}\left(\frac{\delta_L}{\Delta} + \frac{\delta_C}{\beta}\right)c_0,
\en
where $B = \beta\frac{\delta_L}{\Delta} + \delta_C$, $C = \frac{\epsilon}{\Delta} - 2\beta - 1 + \frac{\delta_L^2}{4\Delta^2}$. As in the previous cases, below we exhibit the first coefficients, considering $c_0=1$:
\begin{align}
c_2&=\frac{1}{2\beta+1}\left[\frac{M\beta}{\bar{\omega}}\Omega\left(\Omega + \frac{\eta_L}{\beta M\bar{\omega}}\right)-\frac{C}{2}\right] \ , \nonumber\\
c_3&= \frac{1}{6(\beta+1)\sqrt{M\bar{\omega}}}\left\{\frac{1}{2\beta+1}\left[\frac{4\eta_L}{\bar{\omega}}+2M\beta\Omega\right] \left[\frac{M\beta}{\bar{\omega}}\Omega\left(\Omega + \frac{\eta_L}{M\bar{\omega}}\right) - \frac{C}{2}\right]+M(2-C)\Omega\right\} \ ,
\end{align}
with $\Omega = \frac{\eta_L}{M\bar{\omega}}+\frac{\eta_C}{\beta}$. To obtain a polynomial form of the function, we must impose two conditions on the coefficients similar to \eqref{cond1} and \eqref{cond2}. The condition $C=2n$, provides the energy spectrum, 
\bn
\label{E59}
E_{k,m,n} = \pm\left[k^2 +\frac{M^4\omega^2}{\eta^2_L+M^2\omega^2} + 2\eta_C\eta_L - 2M\omega\frac{m}{\alpha} + 2M\bar{\omega}\left(n+\frac{1}{2}+\beta\right)\right]^{\frac{1}{2}} \ ,
\en
being $\bar{\omega}=\frac1M\sqrt{\eta_L^2+M^2\omega^2}$. We can see that \eqref{E59} does not change by the discrete symmetry $\eta_C\to-\eta_C$ together with $\eta_L\to-\eta_L$.

From this result we can see that the magnetic field and both terms in the scalar potential interfere significantly in the energy spectrum. For the case were the function $H(x)$ becomes a Heun polynomial of first degree,
\bn
\label{H-sol}
H_{1,m}(r) = 1 + M\left[\frac{\eta_L}{M\bar{\omega}_{1,m}} + \frac{\eta_C}{\beta}\right]r \ ,
\en
the energy of the particle is 
\bn
\label{E-F}
E_{k,m,1} = \pm\left[k^2 + \frac{M^4\omega_{1,m}^2}{\eta^2_L+M^2\omega_{1,m}^2} + 2\eta_C\eta_L - \frac{2m}{\alpha}M\omega_{1,m}+ 2M\bar{\omega}_{1,m}\left(\beta + \frac{3}{2}\right)\right]^{\frac{1}{2}} \ .
\en
The condition $c_2=0$  imposes that the parameter $\bar{\omega}_{1,m}$ satisfies the following third degree algebraic equation,
\bn
\label{W-F}
\bar{\omega}_{1,m}^3 - \frac{M\eta_C^2}{\beta}\bar{\omega}_{1,m}^2 - \frac{(2\beta+1)\eta_C\eta_L}{\beta}\bar{\omega}_{1,m} - \frac{(\beta+1)\eta_L^2}{M} = 0 \ .
\en
Although the above equation presents at least one real solution, we decide do not reproduce it here because the latter is a very long expression. We also would like to emphasize that the energy level \eqref{E-F}, has been obtained assuming that the parameters $\eta_C$ and $\eta_L$ are different from zero, $C=2$ and $c_2=0$. This is the lowest energy level that the analytical method adopted can provide. We can see that this energy level is completely different from the corresponding one obtained from the Ladau levels given by \eqref{L-L}.

\section{Fermionic Fields: Dirac Equation}
\label{fermionfields}
In this section we shall consider the relativistic quantum analysis of a spin$-\frac{1}{2}$ charged particle considering the most general situation, i.e., the fermionic particle in the presence of a uniform magnetic field and scalar potential, given, respectively by \eqref{Mag} and \eqref{potential}.

Let us first introduce the generalization of the Dirac equation to an arbitrary curved spacetime, in the presence of an external electromagnetic potential and a scalar potential, which is given by
\bn\label{diraceq}
\left[i\gamma^\mu(x)\left(\nabla_\mu + i e A_\mu \right) - (M + S(r))\right]\Psi(x)=0 \ .
\en
Here $M$ is the mass of the particle and the covariant derivative for fermion fields is defined as follows
\bn
\nabla_\mu = \partial_\mu + \Gamma_\mu \ ,
\en
together with the spinor affine connection 
\bn\label{saffinconn}
\Gamma_\mu = \frac{1}{4}\gamma^{(a)}\gamma^{(b)}e^\nu_{(a)}\left[\partial_\mu e_{(b)\nu} - \Gamma^\sigma_{\mu\nu}e_{(b)\sigma}\right] \ ,
\en
where $\Gamma^\sigma_{\mu\nu}$ is the Christoffel symbol of the second kind and $e^\mu_{(a)}(x)$ the tetrad basis defined later. Also, $\gamma^{\mu}(x)$ represents the generalized Dirac matrix satisfying the Clifford algebra
\bn
\left\{\gamma^{\mu}(x),\gamma^{\nu}(x)\right\} = 2g^{\mu\nu}(x) \ ,
\en
and defined in terms of a set of tetrad fields $e^\mu_{(a)}(x)$ and constant Dirac matrices, $\gamma_{(a)}$, as
\bn\label{gengamma}
\gamma^{\mu}(x) = e^\mu_{(a)}(x)\gamma_{(a)} \ .
\en
The tetrad fields, $e^\mu_{(a)}(x)$, satisfy the relation 
\bn
\eta^{(a)(b)}e^\mu_{(a)}(x)e^\nu_{(b)}(x) = g^{\mu\nu} \ .
\en
Here Greek letters are used for tensor indices and Latin letters (in parenthesis) for tetrad indices. The matrices $\gamma^{(a)}$ are the standard flat spacetime Dirac matrices. In this paper we shall use the definition of such matrices, as given below:
\bn\label{gamma}
\gamma^{(0)} = 
\left[ \begin{array}{cc}
\textbf{1} & 0 \\
0 & -\textbf{1}  
\end{array} \right], \quad {\rm and} \quad
\gamma^{(a)} = 
\left[ \begin{array}{cc}
0             & \sigma^{(a)} \\
-\sigma^{(a)} & 0  
\end{array} \right] \ ,
\en
where $\sigma^{(a)}$ are Pauli matrices and $\textbf{1}$ the identity $2\times2$ matrix.

In order to consider the Dirac equation in the spacetime of a cosmic string, let us consider the tetrad $e^\mu_{(a)}$ as:
\bn\label{tetrad}
e^\mu_{(a)} = 
\left[ \begin{array}{cccc}
1 & 0          & 0                            & 0 \\
0 & \cos{\phi} & -\frac{\sin{\phi}}{\alpha r} & 0 \\
0 & \sin{\phi} & \frac{\cos{\phi}}{\alpha r}  & 0 \\
0 & 0          & 0                            & 1
\end{array} \right] \ .
\en
By using equations \eqref{gengamma}, \eqref{gamma} and \eqref{tetrad}, we obtain the following expressions for the generalized Dirac matrices $\gamma^\mu(x)$:
\begin{align}\label{gendiracmat}
\gamma^0 = \gamma^{(0)}, \quad {\rm and} \quad
\gamma^i = \left[ \begin{array}{cc}
0 & \sigma^i \\
-\sigma^i & 0  
\end{array} \right] \ .
\end{align}
Here, $\sigma^\mu$ represents the following set of modified Pauli matrices:
\begin{align}
\sigma^1 &= \sigma^r = \left[ \begin{array}{cc}
0 & e^{-i\phi} \\
e^{i\phi} & 0  
\end{array} \right],\\
\sigma^2 &= \sigma^\phi = -\frac{i}{\alpha r}\left[ \begin{array}{cc}
0 & e^{-i\phi} \\
-e^{i\phi} & 0  
\end{array} \right],\\
\sigma^3 &= \sigma^z = \sigma^{(3)},
\end{align}
hence $\gamma^3 = \gamma^{(3)}$.

\subsection{Fermionic Radial Equations}\label{fermionradeq}

In this subsection, we shall investigate Dirac's equation \eqref{diraceq}, decoupling the differential equations for the four components of the Dirac spinor, through a series of symmetry considerations. As in previous sections, the presence of the magnetic field and the scalar potential is considered, just as described in section \ref{sec2}.

From equations \eqref{saffinconn}, \eqref{tetrad} and \eqref{gendiracmat}, it is possible to evaluate the spinor connection
\bn
\Gamma_\mu = \left(0,\ 0, \ \Gamma_\phi, \ 0\right), \quad {\rm with} \quad \Gamma_\phi = \frac{i(1-\alpha)}{2}\Sigma^{(3)}, \quad {\rm and}\quad \Sigma^{(3)} = \left[ \begin{array}{cc}
\sigma^{(3)} & 0 \\
0 & \sigma^{(3)}  
\end{array} \right].
\en
The Hamiltonian associated with \eqref{diraceq} reads
\bn
{\cal H} = -i\gamma_{(0)}\left[\gamma^r\partial_r + \gamma^\phi\left(\partial_\phi + \frac{i(1-\alpha)}{2}\Sigma^{(3)} + ieA_\phi(r)\right) + \gamma^{(3)}\partial_z + i(M+S(r))\right] \ .
\en
This operator satisfies the commutation relations $[{\cal H},\hat{p}_3]=[{\cal H},\hat{J}_3]=0$, where $\hat{p}_3 = -i\partial_z$, $\hat{J}_3 = \hat{L}_3 + \hat{S}_3 = -i\partial_\phi + \frac{1}{2}\Sigma^{(3)}$, and the following linear eigenvalue equations hold
\begin{align}
{\cal H} \Psi &= E\Psi \ ,\\
\hat{p}_3\Psi &= k\Psi\ ,\\
\hat{J}_3\Psi &= -i\partial_\phi\Psi + \frac{1}{2}\Sigma^{(3)}\Psi = j\Psi \ ,
\end{align}
where $j = m + \frac{1}{2}$, $m = 0, \pm 1, \pm 2, \ldots$, and $k\in [-\infty, \infty]$. 

We shall adopt the following ansatz for the wave function:
\bn\label{ans-ferm}
\Psi(t,r,\phi,z) = \frac{1}{\sqrt{r}}e^{-iEt + im\phi + ikz}\left[ \begin{array}{c}
\varphi_{+}(r) \\
-i\varphi_{-}(r)e^{i\phi} \\
\chi_{+}(r) \\
i\chi_{-}(r)e^{i\phi}
\end{array} \right] \ ,
\en
and the radial functions obey the following coupled differential equations:
\begin{align}
\chi_{-}'+ \left(\frac{j}{\alpha r} - M\omega r\right)\chi_{-}+ k\chi_{+}+ \left[M -E + S(r)\right]\varphi_{+} &= 0 \ ,\\
\chi_{+}'- \left(\frac{j}{\alpha r} -M\omega r\right)\chi_{+}    + k\chi_{-}    + \left[M - E + S(r)\right]\varphi_{-} &= 0 \ ,\\
\varphi_{-}' + \left(\frac{j}{\alpha r}- M\omega r\right)\varphi_{-} - k\varphi_{+} + \left[M + E + S(r)\right]\chi_{+}  &= 0 \ ,\\
\varphi_{+}'- \left(\frac{j}{\alpha r}-M\omega r\right)\varphi_{+} - k\varphi_{-} + \left[M + E + S(r)\right]\chi_{-} &= 0 \ .
\end{align}
Those equations exhibit the discrete symmetry
\begin{align}\label{symmetryrad}
\chi_{+}    = \lambda\varphi_{+} \quad {\rm and}\quad \varphi_{-} = \lambda\chi_{-} \ , \quad {\rm with} \quad \lambda = \frac{E + s\sqrt{E^2-k^2}}{k} \quad {\rm and} \quad s=\pm 1 \ .
\end{align}
We have also introduced the cyclotron frequency as $\omega=\frac{eB_0}{2M}$. Applying the symmetry argument expressed in \eqref{symmetryrad}, we are able to reduce the number of equations to the only two given below:
\begin{align}
\left[\frac{d^2}{dr^2} - \frac{j(j-\alpha)}{\alpha^2 r^2} + E^2 - k^2 + 2M\omega\left(\frac{j}{\alpha} + \frac{1}{2}\right) - M^2\omega^2r^2 - (M + S(r))^2\right]\varphi + \frac{dS(r)}{dr}\chi &= 0 \ ,\label{dirac1}\\
\left[\frac{d^2}{dr^2} - \frac{j(j+\alpha)}{\alpha^2 r^2}+ E^2 - k^2 + 2M\omega\left(\frac{j}{\alpha} - \frac{1}{2}\right)  - M^2\omega^2r^2 - (M + S(r))^2\right]\chi + \frac{dS(r)}{dr}\varphi &= 0 \ ,\label{dirac2}
\end{align}
where we have used $\varphi_{+} \equiv \varphi$, and $\chi_{-} \equiv \chi$ for simplicity. 

Equations \eqref{dirac1} and \eqref{dirac2} are coupled second order ordinary differential equations. We notice that this pair of equations can be written in the following way
\bn
\hat{K}\left( \begin{array}{c}
\varphi \\
\chi  
\end{array} \right)=0 \ ,
\en
where $\hat{K}$ is the $2\times2$ matrix operator below,
\begin{eqnarray}
\hat{K} = D(r)\textbf{1} + A(r)\sigma^{(3)} + S'(r)\sigma^{(1)} \ ,
\end{eqnarray}
with
\begin{align}
D(r) &= \frac{d^2}{dr^2}- \frac{j^2}{\alpha^2 r^2}+E^2 - k^2 + 2M\omega \frac{j}{\alpha} - M^2\omega^2 r^2 - (M + S(r))^2 \ ,\\
A(r) &= \frac{j}{\alpha r^2} + M\omega \ ,\\
S'(r) &= -\frac{\eta_C}{r^2} + \eta_L \ .
\end{align}
The operator $\hat{K}$ can be diagonalized by using a $2\times 2$ orthogonal matrix $\hat{R}$ as shown,
\bn
\hat{R}\hat{K}\hat{R}^{-1}\hat{R}\left( \begin{array}{c}
\varphi \\
\chi  
\end{array} \right)=0, \quad {\rm with}\quad \hat{R} = \left[ \begin{array}{cc}
\cos(\theta/2) & \sin(\theta/2) \\
-\sin(\theta/2) & \cos(\theta/2)  
\end{array} \right].
\en
This process allows us to write the new operator, named $\hat{K}_\theta$, by
\bn
\hat{K}_\theta = D(r)\textbf{1} + \left(\frac{j}{\alpha r^2} + M\omega\right)\left[ \begin{array}{cc}
\cos{\theta} & -\sin{\theta} \\
-\sin{\theta} & -\cos{\theta}  
\end{array} \right] + \left(\eta_L - \frac{\eta_C}{r^2}\right)\left[ \begin{array}{cc}
\sin{\theta} & \cos{\theta} \\
\cos{\theta} & -\sin{\theta}\end{array} \right] \ .
\en
In conclusion, this differential operator is diagonal if both conditions below are simultaneously fulfilled:
\bn\label{diag1}
\frac{j}{\alpha}\tan{\theta} = -\eta_C \quad {\rm and} \quad M\omega \tan{\theta} = \eta_L \ .
\en
This analysis produces the following set of relations
\begin{align}
\omega &= -\frac{j \eta_L}{\alpha M \eta_C} \ ,\label{omega}\\
\sin{\theta} &= - \frac{\alpha\eta_C}{\sqrt{j^2 + \alpha^2 \eta_C^2}} \ ,\label{diag3}\\
\cos{\theta} &= \frac{j}{\sqrt{j^2 + \alpha^2 \eta_C^2}} \ .\label{diag4}
\end{align}
This means that in order to decouple \eqref{dirac1} and \eqref{dirac2}, one must impose the condition \eqref{omega}. Accepting this condition the analysis of the quantum motion becomes enormously simplified. The two fourth order radial differential equations obtained in the general situation, reduce themselves to two second order differential equations. Although this condition provides a much simpler system, in principle the solutions for the wave equations and energy spectra may not be compatible. Fortunately, as we shall see later, this situation does not occur in this analysis. Now returning to \eqref{omega}, we see that the magnetic field strength, the mass of the particle and the scalar coupling constants must be related, and, on the other hand, this relations depends on the total angular momentum quantum number, $j$. In order to fulfill this condition, we may choose that the parameters $\omega$ or $\eta_L$, must be adjusted to a specific value if we are inclined to find exact solutions to the radial equations. As we did in the bosonic case, and accepting the condition \eqref{omega}, it is possible to rewrite the previous differential equations in a Schr\"odinger form, as follows:
\begin{align}
\frac{d^2\varphi}{dr^2} + \left[E^2 - M^2 - k^2 + 2M\omega\frac{j}{\alpha} - 2\eta_C\eta_L - \gamma\frac{\eta_L}{\eta_C} - \frac{\gamma(\gamma-1)}{r^2} - \frac{2M\eta_C}{r} -2M\eta_L r - \gamma^2\frac{\eta_L^2}{\eta_C^2} r^2\right]\varphi &= 0 \ ,\label{diracradial1}\\
\frac{d^2\chi}{dr^2}    + \left[E^2 - M^2 - k^2 + 2M\omega\frac{j}{\alpha} - 2\eta_C\eta_L + \gamma\frac{\eta_L}{\eta_C} - \frac{\gamma(\gamma+1)}{r^2} - \frac{2M\eta_C}{r} -2M\eta_L r - \gamma^2\frac{\eta_L^2}{\eta_C^2} r^2\right]\chi &= 0 \ ,\label{diracradial2}
\end{align}
where $\gamma = \frac{1}{\alpha}\sqrt{j^2 + \alpha^2\eta_C^2}$. Moreover, we can write these equations as BCH ones given by \eqref{BCHschr}. Adopting $\eta_C$ as a positive constant, and defining a new dimensionless variable
\bn\label{variabledirac}
x = \sqrt{\gamma \frac{|\eta_L|}{\eta_C}} \ r \ ,
\en
we have
\begin{align}
\varphi''(x) + \left[\frac{\zeta_{k,j}\eta_C}{\gamma |\eta_L|} - \frac{\eta_L}{|\eta_L|} - \frac{\gamma(\gamma - 1)}{x^2} + \frac{a}{x} + bx - x^2\right]\varphi(x) &= 0 \ ,\label{diracdiff1}\\
\chi''(x)    + \left[\frac{\zeta_{k,j}\eta_C}{\gamma |\eta_L|} + \frac{\eta_L}{|\eta_L|} - \frac{\gamma(\gamma + 1)}{x^2} + \frac{a}{x} + bx - x^2\right]\chi(x)    &= 0 \ ,\label{diracdiff2}
\end{align}
where
\begin{align}
\zeta_{k,j} &= E^2 - M^2 - k^2 + 2M\omega\frac{j}{\alpha} - 2\eta_C\eta_L \ ,\\
a &= -2M\eta_C\sqrt{\frac{\eta_C}{\gamma|\eta_L|}} \ ,\\
b &= \frac{a\eta_L}{\gamma |\eta_L|}.
\end{align}
We stop at this point and leave the obtainment of analytical solutions to \eqref{diracdiff1} and \eqref{diracdiff2} for section \ref{soldirac}.

\section{Analytical Solutions for the Dirac Equation of Motion}
\label{soldirac}

Following in the same direction as we did to obtain solutions of the Klein-Gordon equation, it is convenient to analyze the asymptotic behavior of the solutions of the radial differential equations \eqref{diracradial1} and \eqref{diracradial2} for $r\longrightarrow 0$ and $r \longrightarrow \infty$. Through this analysis, we verify that the radial components of the Dirac spinor can be expressed as
\begin{align}
\varphi(x) &= x^\gamma e^{-\frac{1}{2}x\left(x - b\right)}F(x) \ ,\\
\chi(x) &= x^{\gamma+1} e^{-\frac{1}{2}x\left(x - b\right)}H(x) \ .
\end{align}
Inserting this ansatz into \eqref{diracdiff1} and \eqref{diracdiff2}, we obtain
\begin{eqnarray}
xF''(x)&+&\left(2\gamma+bx - 2x^2\right)F'(x) + \left[\left(1 + \frac{\eta_L}{|\eta_L|}\right)a \right.\nonumber\\
&+&\left. \left(\frac{\zeta_{k,j}\eta_C}{\gamma |\eta_L|} + \frac{b^2}{4}- 2(\gamma + 1) + 1 -\frac{\eta_L}{|\eta_L|} \right)x\right]F(x) = 0 \ ,\label{bchdirac1}\\
xH''(x)&+&\left(2(\gamma+1)+bx - 2x^2\right)H'(x)+\left[\left(1+\frac{(\gamma+1)\eta_L}{\gamma|\eta_L|}\right)a\right.\nonumber\\
&+&\left.\left(\frac{\zeta_{k,j}\eta_C}{\gamma |\eta_L|} + \frac{b^2}{4}- 2(\gamma + 1) - 1 +\frac{\eta_L}{|\eta_L|}\right)x\right]H(x) = 0 \ .\label{bchdirac2}
\end{eqnarray}
These are BCH equations with solutions
\begin{align}
F(x) = H_B\left(2\gamma - 1,\quad b,\quad \frac{\zeta_{k,j}\eta_C}{\gamma |\eta_L|} + \frac{b^2}{4} - \frac{\eta_L}{|\eta_L|}, \quad 2a; \quad -x\right) \ ,\\
H(x) = H_B\left(2\gamma + 1,\quad b,\quad \frac{\zeta_{k,j}\eta_C}{\gamma |\eta_L|} + \frac{b^2}{4} + \frac{\eta_L}{|\eta_L|}, \quad 2a; \quad -x\right) \ .
\end{align}

It is useful to follow the analyses of both equations \eqref{bchdirac1} and \eqref{bchdirac2} simultaneously. To accomplish this we may define the constants $A$, $B$ and $D$, and the function $G$ representing $F$ or $H$ as shown below:
\begin{align}
A = \left\{ \begin{array}{c}
\gamma, \quad {\rm if} \quad G=F \\
\gamma + 1, \quad {\rm if} \quad G=H
\end{array} \right. \quad 
B = \left\{ \begin{array}{c}
1 + \frac{\eta_L}{|\eta_L|}, \quad {\rm if} \quad G=F \\
1+\frac{(\gamma+1)\eta_L}{\gamma|\eta_L|}, \quad {\rm if} \quad G=H
\end{array} \right. \\ {\rm and}\quad 
D = \left\{ \begin{array}{c}
\frac{\zeta_{k,j}\eta_C}{\gamma |\eta_L|} + \frac{b^2}{4}- 2(\gamma + 1) + 1 -\frac{\eta_L}{|\eta_L|}, \quad {\rm if} \quad G=F \\
\frac{\zeta_{k,j}\eta_C}{\gamma |\eta_L|} + \frac{b^2}{4}- 2(\gamma + 1) - 1 +\frac{\eta_L}{|\eta_L|}, \quad {\rm if} \quad G=H
\end{array} \right. \ .
\end{align}
By making use of theses definitions, it is possible to write the two differential equations in just one single form,
\bn
xG'' + \left(2A + bx - 2x^2\right)G' + \left(Ba + Dx\right)G = 0 \ .\label{G}
\en
Here we shall use Frobenius method to find series expansions to this solution: 
\bn
G(x)= \sum_n C_n x^n \ .
\en
Substituting the above series into \eqref{G}, we obtain the recurrence relation
\begin{align}
C_{n+2} &= \frac{-1}{(n+2)(n+2A + 1)}\left\{\left[Ba + b(n+1)\right]C_{n+1} + \left(D-2n\right)C_n\right\} \\
{\rm and}\quad C_1 &= -\frac{Ba}{2A}C_0 \ .
\end{align}
With this relation we can construct the first three coefficients of the expansion. Assuming $C_0 = 1$ they are:
\begin{align}
C_2 &= \frac{1}{2(2A + 1)}\left[\frac{Ba}{2A}\left(Ba + b\right) - D\right] \ ,\\
C_3 &= \frac{1}{6(A + 1)}\left\{\frac{Ba + 2b}{2(2A+1)}\left[D - \frac{Ba}{2A}\left(Ba + b\right)\right] + \frac{Ba}{2A}(D-2)\right\} \ .
\end{align}
At this point we know that by breaking the series expansion of the BCH function into a Heun polynomial of degree $n$, we obtain an analytical solution to the radial equations. This can be implemented by the imposition of two conditions on the coefficients in a way similar to what we did in the bosonic analyses. These conditions are:
\bn
C_{n+1} = 0 \quad {\rm and}\quad D=2n \ , \quad {\rm with}\quad n = 1, 2, ...
\en
From the condition $D = 2n$, it is possible to obtain a formal expression to the energy. We should be able to reproduce the same energy eigenvalue independently of the choice we make for the definition of the constant $D$ (either for $G=F$ and $G=H$). For this to occur, we need to impose $\eta_L >0$, otherwise, there are no eigenvalues associated with the Heun polynomials which satisfy the condition $D = 2n$ for both components. After adopting this restriction, we obtain:
\bn\label{enspecdirac}
E_{k, j, n} = \pm\left\{M^2 - k^2 - 2M\omega\frac{j}{\alpha} + 2\eta_C{\eta_L}_n + {\eta_L}_n\left[\frac{2\gamma(\gamma + n + 1)}{\eta_C} - \frac{M^2\eta_C^2}{\gamma^2{\eta_L}_n}\right]\right\}^{\frac{1}{2}}.
\en
If we assume the solution as a Heun polynomial of degree $n=1$, the equation $C_2 = 0$ provides that the parameter $\eta_L$ will depend on the total angular momentum quantum number $j$. So for this case we have 
\bn\label{paramdirac}
{\eta_L}_{1,j} = \frac{2M^2\eta_C^3(2\gamma+1)}{\gamma^3} \ .
\en
The above value together with \eqref{enspecdirac} provide the energy spectrum of the particle.
\section{Concluding Remarks}
\label{concl}

In this paper we have investigated the relativistic quantum motion of charged spin-$0$ and spin$-\frac{1}{2}$ particles in a cosmic string spacetime, in the presence of a uniform magnetic field parallel to the string and scalar potentials whose centers are on the string. The complete analysis is twofold: it is applied to bosonic particles in the first part followed by the fermionic case.

For the bosonic case we obtained from the Klein-Gordon equation, a radial differential equation \eqref{radialgen} by adopting the ansatz \eqref{Phi} for the wave-function. The former is analyzed considering different values for the physical parameters. Specifically for the most general case investigated in subsection \ref{sub4}, we were able to express the corresponding differential equation in terms of the biconfluent Heun differential equation. Because we are interested in obtaining an explicit expression for the energy spectrum, we searched for polynomial solutions for this radial equation. So, in order to obtain these polynomials, we used the Frobenius method and a three-term recurrence relation for the coefficients of the expansion was obtained. Finally, by imposing specific conditions on these coefficients, polynomials solutions are found. Admitting a first order polynomial solution, the energy found was given by \eqref{E-F} under the condition that the cyclotron frequency is a real solution of \eqref{W-F}. Because this solution is a long expression, we decided to not reproduce it in the paper.

For the fermionic case, we observed that following the particular choice of the ansatz for the wave-function \eqref{ans-ferm}, a set of four radial differential equations is obtained from the Dirac equation. Moreover, by adopting the discrete symmetry given by \eqref{symmetryrad}, we were able to reduce the system to only two coupled differential equations. A direct attempt to decouple these equations would result in a much more complex system of fourth-order differential equations. Instead, through the relations \eqref{diag1}-\eqref{diag4}, which require that both scalar potentials be considered (this is particularly clear in \eqref{diag1}), it is possible to greatly simplify this set of differential equations into biconfluent Heun equations. As happened in the bosonic case, it is also useful to simplify the BCH solutions into Heun polynomials in order to obtain the energy spectrum of the quantum particle. This provides an additional algebraic relation between the parameters of the system.

In conclusion, we notice that a simpler analysis of the motion of the fermionic field can be obtained, provided that two conditions involving the parameters of the system be satisfied, namely \eqref{omega} and \eqref{paramdirac} for the case where the Heun polynomial is of the first degree. Because we have accepted $\eta_C>0$, to obtain a physically relevant result for the energy spectrum, we have also to assume that $\eta_L>0$. As a consequence, in order to \eqref{omega} be satisfied, for a magnetic field along the positive direction of $z-$axis, i.e., $\vec{B}=B_0{\hat{k}}$, the value of the total angular momentum quantum number, $j$,  must be negative. Having chosen the opposite direction to the magnetic field only positive values for $j$ would be accepted. We conclude that the general case must be dealt with by means of two different analyses: for a magnetic field along the direction $\hat{k}$ a negative value of the total angular quantum number is required; moreover, for the magnetic field in the opposite direction, a positive value is necessary. The fact that the sign of the quantum number $j$ is conditioned to the direction of the magnetic field does not represent a strong restriction in our result, since the energy spectrum obtained in \eqref{enspecdirac} depends on the product between $\omega$ and $j$ which is always negative.

As we have seen, closed analyses involving Heun biconfluent equations can only be carried out by imposing a specific Heun polynomial of degree $n$ for the solutions of the field equations. The reason for this difficulty is that, for the complete biconfluent Heun function, there is no general expression for its asymptotic behavior for large values of the argument.\footnote{For specific values of the parameters, the BCH function can be expressed as the confluent hypergeometric function.} In this present analysis, for bosonic and fermionic particles, we have illustrated how relevant physical informations about the systems can be obtained when the BCH function is expressed as a Heun polynomial of degree $n=1$ (of course this analysis can be extended for any polynomial of degree greater than unity). As consequence of this approach, some physical parameter must be fixed by the order of the polynomial, $n$, and also by the angular quantum numbers. The analysis of a BCH differential equation was also developed by Ver\c cin \cite{Ver} and Ver\c cin {\it at al} \cite{Ver1}, investigating the quantum planar motion of two anions with Coulomb interaction between them and in the presence of an external uniform magnetic field. In the first paper the author found exactly the energy spectrum, by imposing that the magnitude of the magnetic field obeys a closed relation with the angular quantum number. In the second, the energy spectrum was evaluated numerically.

The Coulomb-type scalar potential presented in this paper, can be associated with the presence of an electrostatic self-interaction on the charged particle induced by the non-trivial topology of the conical spacetime. As it was shown numerically in \cite{Smith}, this self-interaction is positive for $\alpha=1-4\mu<1$, which corresponds positive linear mass density of the string $\mu$; an attractive self-interaction takes place for a negative value of $\mu$. As to the linear scalar potential, it can be associated with an extra cylindrical harmonic oscillator acting on the particle, whose classical angular frequency, $\omega_c$, being given by $\eta_L/M$. Moreover, the linear potential, whose original motivation comes from the confining model for quarks adopted in \cite{Chritc,Chritc1}, was basically included in this paper to make the present analysis as general as possible.

Due to the magnetic field and the linear scalar potential, only bound states solutions can be presented in the systems considered. By choosing specific relations between the scalar potentials strength and the magnetic filed, the analysis of the energy-spectra become drastically simplified. Accepting these relations we were able to obtain explicitly the spectra associated with both, bosonic and fermionic charged particles; moreover, from these results we can also observe that, although the quantum particles move in a region of spacetime that is locally flat, the correspond energy spectra of the particles depend on the global properties of the spacetime through the parameter $\alpha$. In fact these energy levels depend on the the orbital angular quantum number for bosonic particles and total angular quantum number for fermionic ones in the combination $m/\alpha$ and $j/\alpha$, respectively. These facts are consequence of the planar angle deficit associated with the conical structure of the two surface orthogonal to the cosmic string. 
 
Before to finish this paper we would like to mention two potential applications of our computation. They are in Condensed Matter and Astrophysics. As to condensed matter physics, it is well known that linear defect in elastic solid named disclination can be dealt with the same geometric approach as cosmic strings \cite{Katanev}. In fact the analysis developed here constitutes a relativistic extension, with some new ingredients, of the previous one presented in \cite{Fur} to study the quantum mechanical motion of charged particles under the influence of an external uniform magnetic field in an elastic medium with disclination. Moreover, specifically to the analysis in section \ref{soldirac}, of the quantum mechanic motion of a fermionic charged particle, it can be adapted to study graphitic cones under the influence of uniform magnetic field within the framework of long-wavelength Dirac-like model for electronic states in graphene (for a review see \cite{Cast}). With respect to astrophysical application, this analysis may also be useful to understand the quantum motion of charged particle in the cosmos under influence of galactic magnetic field considering the presence of a cosmic string\footnote{An estimated strength for our galactic magnetic field can be found in \cite{Parker}. This value is of order $3\times10^{-6}$ gauss.}. Summarizing, we can say that, by the results found in this paper, the presence of a cosmic string produces significant modifications on the energy spectra associated with bosonic and fermionic charged particles the presence of uniform magnetic fields. 
\appendix
\section{Heun's Differential Equation}
\label{app}

The Heun's equation is the generic differential equation with four regular singular points at $0$, $1$, $a$ and $\infty$. The standard natural form of Heun's equation reads \cite{Heun},
\bn
y''(z) + \left(\frac{\lambda}{z} + \frac{\mu}{z-1} + \frac{\varepsilon}{z-a}\right)y'(z) + \left(\frac{\alpha\sigma z - q}{z(z-1)(z-a)}\right)y(z) = 0 \ .
\en
The parameter $\varepsilon$ is expressed in terms of the other ones by $\varepsilon = \alpha + \sigma + 1 - \lambda -\mu$. Four confluent equations arise from the general Heun equation by means of different confluence processes. Of particular interest to our investigation is the biconfluent equation (BCH), with a regular singular point at $z = 0$ and irregular singular point at $z=\infty$. The canonical form of the BCH is 
\bn
\label{Heq}
xu''(x) + (1+\alpha -\sigma x-2x^2)u'(x) +\left\{\left(\lambda - \alpha - 2\right)x - \frac{1}{2}\left[\mu + \sigma\left(1+\alpha\right)\right]\right\}u(x)=0 \ ,
\en
where $\alpha$, $\sigma$, $\lambda$ and $\mu$ are arbitrary parameters. By means of the transformation
\bn
u(x) = x^{-\frac{1+\alpha}{2}}e^{\frac{\sigma x + x^2}{2}}v(x) \ ,
\en
the BCH equation takes on the Schr\"odinger form
\bn\label{BCHschr}
v''(x)+ \left[Ax^2 + Bx + C + \frac{D}{x} + \frac{E}{x^2}\right]v(x) = 0 \ ,
\en
with $A = -1$; $B = -\sigma$; $C = \lambda - \frac{\sigma^2}{4}$; $D =-\frac{\mu}{2}$; $E = \frac{1-\alpha^2}{4}$. 

The biconfluent Heun polynomial, $H_B(\alpha, \sigma, \lambda, \mu; x)$, can be obtained as a Frobenius solution to BCH equation computed as a power series expansion around the origin,
\bn
H_B=\sum_{n=0}^\infty A_n \ x^n \ ,
\en
Substituting the above expansion into \eqref{Heq}, a three-term recurrence relation for the coefficients is obtained:
\bn
A_{m+2}=\frac{1}{(n+2)(n+2+\alpha)}\left\{\left[\frac{\mu}{2} + \sigma\left(n+1 + \frac{\alpha+1}{2}\right)\right]C_{n+1} + \left(2n-\lambda + \alpha + 2\right)C_n\right\},
\en
for $m\geq 1$ and
\bn
A_1=\frac{1}{2}\left[\sigma + \frac{\mu}{1+\alpha}\right]A_0 \ .
\en
The roots of $A_{n+1}$ are the eigenvalues corresponding to these particular Heun polynomials. In this case, all subsequent coefficients cancel and the series results in a polynomial form of degree $n$ for $H_B$. For this case we must have,
\bn
\lambda - \alpha - 2 = 2n \ , \quad {\rm for} \quad n=0,1,2,3... \quad {\rm and} \quad A_{n+1} = 0 \ .
\en

Unfortunately, there is no closed expression for the asymptotic behavior of the biconfluent Heun function in the general case for large values of its argument. Consequently, the conditions imposed on its parameters in order to provide a polynomial form for them, can only be obtained by analyzing each coefficient separately as we did along this paper. 

The biconfluent Heun function can be expressed in terms of the Whittaker $M$ function, or confluent hypergeometric function, for specific values of the parameters, as shown  below:
\bn
H_B(\alpha, 0, \lambda, 0; x) = \frac{1}{x^{\frac{\alpha}{2} + 1}}e^{\frac{x^2}{2}}M\left(\frac{\lambda}{4}, \frac{\alpha}{4}; x^2\right)={}_1F_1\left(\frac\alpha4-\frac\lambda4+\frac12,\frac\alpha2+1; \ x^2\right) \ .
\en

\section*{Acknowledgment} 

The authors thank Conselho Nacional de Desenvolvimento Cient\'\i fico e Tecnol\'ogico (CNPq.).


\begin{thebibliography}{100}
\bibitem{AS} J. Audretsch and G. Sch\"afer, {\it Gen. Rel. Grav.} {\bf 9}, 243 (1978); J. Audretsch and G. Sch\"afer, {\it Gen. Rel. Grav.} {\bf 9}, 489
(1978). 
\bibitem{Parker} L. Parker, {\it Phys. Rev. Lett.} {\bf 44}, 1559 (1980).
\bibitem{Parker1} L. Parker and L. O. Pimentel, {\it Phys. Rev.} D {\bf 25}, 3180 (1982).
\bibitem{GB} G. A. Marques and V. B. Bezerra, {\it Class. Quantum Grav.} {\bf 19}, 985 (2002).
\bibitem{GB1} G. A. Marques and V. B. Bezerra, {\it Phys. Rev.} D {\bf 66}, 105011 (2002).
\bibitem{Kibble} T. W. B. Kibble, {\it J. Phys.} A {\bf 9}, 183 (1976).
\bibitem{Vilenkin} A. Vilenkin, {\it Phy. Rep.} {\bf 121}, 263 (1985).
\bibitem{Linet1} B. Linet, Phys. Rev. D {\bf 33}, 1833 (1986).
\bibitem{Smith} A. G. Smith, in {\it Proceedings of Symposium on the Formation and Evolution of Cosmic Strings}, ed. by G. W. Gibbons, S. W. Hawking and T. Vachaspati (Cambridge University Press, Cambridge, England, 1990).
\bibitem{Furtado} E. R. Bezerra de Mello, V. B. Bezerra, C. Furtado and F. Moraes, {\it Phys. Rev.} D {\bf 51}, 7140 (1995).
\bibitem{Fur} C. Furtado, B. G. C. da Cunha, F. Moraes, E. R. Bezerra de Mello and V. B. Bezerra, {\it Phys. Lett.} A {\bf 195}, 90 (1994).
\bibitem{Mello} E. R. Bezerra de Mello and V. B. Bezerra, {\it J. Math. Phys.} {\bf 36}, 5297 (1995).
\bibitem{DJM} H. G. Dosch, J. H. D. Jansen and V. F. M\"uller, {\it Physical Norvegica} {\bf 5}, 2 (1971).
\bibitem{GBW} G. Soff, B. M\"uller, J. Rafelski and W. Greiner, {\it Z. Naturf.} {\bf 28a}, 1389 (1973).
\bibitem{Nail} M. Bordag and N. Khusnutdinov, {\it Class. Quantum Grav.} {\bf 13}, L41 (1996).
\bibitem{Spi} J. Spinelly, E. R. Bezerra de Mello and V. B. Bezerra, {\it Class. Quantum Grav.} {\bf 18}, 1555 (2001).
\bibitem{Chritc} C. L. Chrichfield, {\it Phys. Rev.} D {\bf 12} 923 (1975).
\bibitem{Chritc1} C. L. Chrichfield, {\it J. Math. Phys.} {\bf 17} 261 (1976).
\bibitem{Su} S. Ru-keng and Z. Yuhong, {\it J. Phys. A: Math. Gen} {\bf 17} 851 (1985).
\bibitem{ALO} A. L. Cavalcanti de Oliveira and E. R. Bezerra de Mello, {\it Class. Quantum Grav.} {\bf 23}, 5249 (2006).
\bibitem{Heun} S. Yu. Slavyanov and W. Lay, {\it Special Functions: A Unified Theory Based in Singularities} (Oxiford Univ. Press, New York, 2000).
\bibitem{Landau} L. D. Landau and E. M. Lifishitz, {\it Course of Theoretical Physics: Quantum Mechanics} (Pergamon Press, 3rd Ed., 1991).
\bibitem{Cooper} F. Cooper, A. Khare and U. Sukhtame, {\it Supersymmetry in Quantum Mechanics} (World Scientific, 2001, Singapore).
\bibitem{Sokolov} A. A. Sokolov and I. M. Ternov, {\it Radiation from Relativistic Electron} (American Institute of Science, 1986).
\bibitem{Abra} M. Abramowitz and I.A. Stegun, {\it Handbook of Mathematical Functions} (Dover, New York, 1972).
\bibitem{Ver} A. Ver\c cin, {\it Phys. Lett.} B {\bf 260}, 120 (1991).
\bibitem{Ver1} J. Myrheim, E. Halvorsen and A. Ver\c cin, {\it Phys. Lett.} B {\bf 278}, 171 (1992).
\bibitem{Katanev} M. O. Katanaev and I. V. Volovich, {\it Ann. Phys.} (NY) {\bf 216}, 1 (1992).
\bibitem{Cast} A.H. Castro Neto, F. Guinea, N.M.R. Peres, K.S. Novoselov, and A.K. Geim, {\it Rev. Mod. Phys.} {\bf 81}, 109 (2009).
\bibitem{Parker} E. N. Parker, {\it Astrophys. J} {\bf 160}, 383 (1970); M. S. Turner, E. N. Parker and T. J. Bogdan, {\it Phys. Rev.} D {\bf 26}, 1296 (1982).
\end{thebibliography}
\end{document}